\documentclass[12pt,preprint]{aastex}

\begin{document}

\title{SOUND-SPEED INVERSION OF THE SUN USING A NONLOCAL STATISTICAL CONVECTION THEORY}

\author{Chunguang Zhang\altaffilmark{1,2}, Licai Deng\altaffilmark{1}, Darun Xiong\altaffilmark{3}, and J\o rgen Christensen-Dalsgaard\altaffilmark{4}}

\altaffiltext{1}{Key Laboratory of Optical Astronomy, National Astronomical Observatories, Chinese Academy of Sciences, Beijing 100012, China; cgzhang@nao.cas.cn}
\altaffiltext{2}{Graduate University of Chinese Academy of Sciences, Beijing 100049, China}
\altaffiltext{3}{Purple Mountain Observatory, Chinese Academy of Sciences, Nanjing 210008, China}
\altaffiltext{4}{Stellar Astrophysics Centre, Department of Physics and Astronomy, Aarhus University, Ny Munkegade 120, DK-8000 Aarhus C, Denmark}

\begin{abstract}
Helioseismic inversions reveal a major discrepancy in sound speed between the Sun and the standard solar model just below the base of solar convection zone. We demonstrate that this discrepancy is caused by the inherent shortcomings of the local mixing-length theory adopted in the standard solar model. Using a self-consistent nonlocal convection theory, we construct an envelope model of the Sun for sound-speed inversion. Our solar model has a very smooth transition from the convective envelope to the radiative interior; and the convective energy flux changes sign crossing the boundaries of the convection zone. It shows evident improvement over the standard solar model, with a significant reduction in the discrepancy in sound speed between the Sun and local convection models.
\end{abstract}


\section{INTRODUCTION}

With the numerous oscillation modes that have been identified with great accuracy and so many well-defined global quantities of the Sun, helioseismology allows us to study the internal structure of the Sun in unprecedented detail. The observed acoustic modes depend predominantly on the sound-speed profile in the solar interior. By inverting solar oscillation frequencies it is shown that the difference of sound speed between the Sun and the standard solar model (SSM) is fairly small, except below the base of the convection zone \citep{Gough96,Basu97}. There is a bump in the sound-speed difference between the Sun and SSM at $r\simeq0.67R$. Here $r$ is the distance to the center and $R$ is the surface radius of the Sun. This sharp feature is generally attributed to the strong composition gradient caused by helium settling. Thus several mechanisms have been introduced to soften the composition gradient or otherwise smooth the sound-speed transition at the base of the convection zone (for a review, see \citealt{JCD11}, and references therein), such as partial mixing, early mass loss, and modification of opacity, composition or temperature profile. However, all these proposed solutions are somewhat ad hoc; some are not even physically realistic.

In fact, a more reasonable explanation of this sound-speed difference may come from a more profound aspect of stellar modeling---the convection theory. In SSM, convection is treated with the mixing-length theory \citep[MLT;][]{BV1958}, which is an oversimplified approach to the fully developed turbulence in stars. One serious deficiency of MLT is that it is a local theory. Thus the properties of the convective turbulence in any place of the fluid field are assumed to depend only on the local structure at that point. This approximation is tolerable in the bulk of the convection zone, but it becomes definitely invalid in the transition zone between the convective envelope and the radiative interior \citep{DX08}.

\section{THE NONLOCAL STATISTICAL CONVECTION THEORY}

Although with apparent deficiencies, MLT is still the most widely used convection theory in stellar modeling. This is mainly because, besides its simplicity, MLT turns out to be very successful in most cases. In solar modeling, the near-surface part of the convection zone is significantly superadiabatic. The entropy jump between the atmosphere and the adiabatic convective region, is a measure of the efficacy of convection, which in MLT is controlled by the mixing-length. This in turn determines the solar radius. The mixing length in MLT can be adjusted to give the appropriate entropy jump to obtain the correct radius. In the interior of the convection zone, the energy transfer by convection is very effective, and the temperature gradient is nearly adiabatic, which hardly depends on a specific convection theory. However, MLT predicts a sudden stop of convective motions at the boundary defined by buoyancy neutrality (i.e., the Schwarzschild criterion). This is not physically realistic since the fluid elements will move past the boundary because of inertia. The local nature of MLT makes it inadequate for the treatment of overshooting.

We have developed a statistical theory of nonlocal convection \citep{Xiong79,Xiong89,DXC06}. This theory is derived rigorously from the basic hydrodynamic equations to construct the dynamic equations of auto- and cross-correlations of turbulent velocity and temperature. The nonlocal effects of convection are represented by the third-order moments. Because of the nonlinearity of the hydrodynamic equations, these moment equations form an unclosed hierarchy, and thus require a suitable closure approximation before they can be solved. Here a gradient-type diffusion approximation, in which the third-order terms are expressed as the gradient of the corresponding second-order terms, is adopted. The theory depends on two dimensionless coefficients $c_1$ and $c_2$ which are related to turbulent dissipation and diffusion, respectively. They can be adjusted to obtain the correct depth of the convection zone and an appropriate extent of the convective overshooting zone. Like the free parameter in MLT, they cannot be determined within the theory, but have to be constrained by observation.

By closing the moment equations at the third-order moments, the nonlocal transport of turbulent energy and momentum are included in our nonlocal formulation of convection. Therefore the convective motions in one place of the fluid field also depend on the properties of the fluid in other regions. In the overshooting zone, nonlocal turbulent diffusion serves as the driving mechanism when both buoyancy breaking and turbulent dissipation decelerate the convective flows \citep{DX08}. Besides our gradient-type approximation, the moment equations can also be solved using other closure schemes (e.g., \citealt{Canuto92}; for a detailed discussion, see \citealt{Grossman96}, and references therein).

Our nonlocal theory is more capable of dealing with the nonlocal effects of convection, though it is more sophisticated and therefore more complicated in applications than MLT is. Nevertheless, the theory has been successful in solving problems of stellar structure, evolution, and oscillations \citep{Xiong86,XD07,XD09}.

\section{THE SOLAR ENVELOPE MODEL}

We use our nonlocal convection theory to construct a new solar model, in the following Model NL. Since a commonly used SSM in helioseismology is Model S of \citet{JCD96}, Model NL has been calibrated to the same basic properties of Model S, including the photospheric radius $R = 6.9599 \times 10^{10}\,\rm{cm}$ and surface luminosity $L_\odot = 3.846 \times 10^{33}\,\rm{erg\ s^{-1}}$. It is calculated with a uniform chemical composition. To compare with Model S, we use the same chemical composition \citep[i.e.,][]{GN93}. We also use the OPAL equation of state \citep{RSI96} and opacities \citep{IR96}. For temperature lower than $6000\,$K, opacity tables of \citet{Ferguson05} are taken as a supplement.

To avoid extra complications such as nuclear reactions and chemical evolution in the core, our calculations are limited to envelope models with the bottom set at $0.3R$. This is well below the bottom of the convection zone, and will not compromise our inversion for the structure of the convective envelope.

Model NL shows major differences from local models near the boundaries of the convection zone, especially in the lower transition region between solar convection zone and radiative interior. Figure \ref{nabla} shows the behavior of the temperature gradient $\nabla = d \ln \it{T}/d \ln \it{P}$, $T$ being temperature and $P$ pressure, in this region for both Model NL and Model S. It is clear that the transition from the convection zone to the radiative interior in Model NL is much smoother than that in Model S. In Model S, $\nabla$ changes abruptly from the adiabatic temperature gradient $\nabla_{\rm{ad}}$ to the radiative temperature gradient $\nabla_{\rm{rad}}$ crossing the lower boundary of the convection zone. However, in Model NL, $\nabla$ is already sub-adiabatic before reaching the boundary, thus there is no abrupt change at the boundary.

\begin{figure}
\center
\includegraphics[scale=.7]{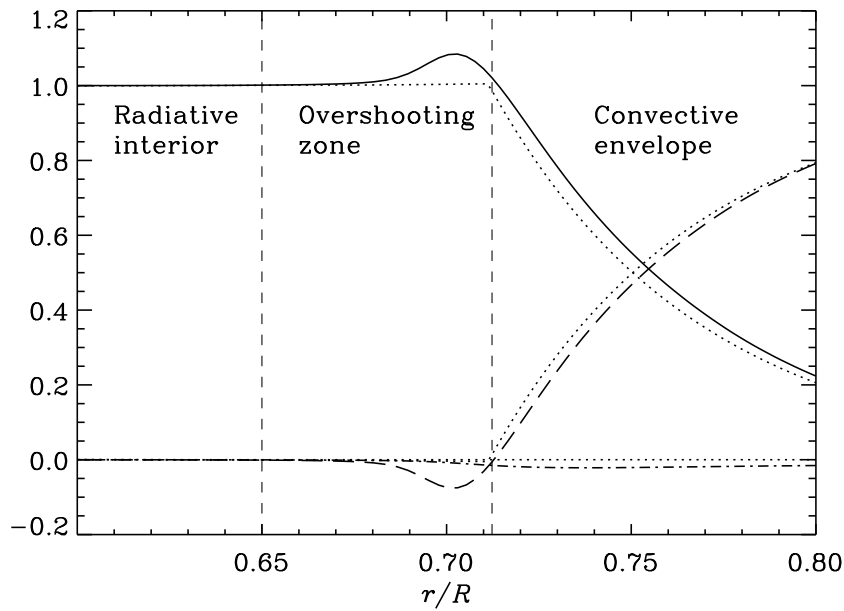}
\caption{Fractional radiative energy flux $F_{\rm r}/F$ (continuous line), convective energy flux $F_{\rm c}/F$ (long dashed line), and turbulent kinetic energy flux $F_{\rm k}/F$ (dash-dotted line) vs. fractional radius $r/R$ for Model NL. The dotted lines show the corresponding fractional fluxes of Model S; and the dashed lines mark the boundaries of the overshooting region in Model NL.}
\label{flux}
\end{figure}

Figure \ref{flux} shows the fractional radiative energy fluxes $F_{\rm r}/F$ and convective energy fluxes $F_{\rm c}/F$ versus fractional radius $r/R$ for both models. In Model S, the convective energy flux $F_{\rm c}>0$ in the convectively unstable zone, and $F_{\rm c}=0$ in the convectively stable zone. In Model NL, $F_{\rm c}$ (or the velocity--temperature correlation) is positive in the convection zone, but becomes negative in the overshooting zone \citep{XD01,DX08}. It is clear in Figure \ref{flux} that the convective flux $F_{\rm c}$ changes sign crossing the lower boundary of the convection zone. In the overshooting zone, $F_{\rm c}$ is negative and the radiative flux $F_{\rm r}$ is larger than the total flux of the Sun. As a result, the temperature in the overshooting zone beneath the convection zone will increase. Thus the sound speed in this region is higher than MLT predicts, which is in accordance with the results of helioseismology.

In Model NL, the change of sign of the velocity-temperature correlation happens at both boundaries of the convection zone. At the upper boundary, this theoretical prediction has been proved by observation. In the classic work of \citet{LNS62}, they reported the reversal of sign of the brightness/temperature-velocity correlation in the solar atmosphere. The same conclusion was drawn by later observations with new technology \citep{SBCCR94}. The properties of the lower overshooting zone can be verified by numerical simulation. Two-dimensional and three-dimensional numerical simulations of compressible convection \citep{HTM86,MZ93} show that in the overshooting zone, the convective energy flux becomes negative. \citet{Kupka99} confirmed this consistency between numerical simulations and the moment equations.

In local convection models, the lower boundary of the convection zone is defined by the Schwarzschild criterion $\nabla=\nabla_{\rm{ad}}$. However, the smooth transition of $\nabla$ in the nonlocal convection model makes it difficult to give such a clear definition. \citet{DX08} argued that a proper definition of the boundary of the convection zone should be the place where the convective energy flux $F_{\rm c}$ (i.e., the correlation of turbulent velocity and temperature) changes sign. The physical meaning of this definition is that the convection zone is the zone of buoyant force driving for convective motion, while the overshooting zone is the dissipation zone against convective motion. Moreover, using this definition, the local and non-local models with the same depth of convection zone will have similar structures.  As illustrated in Figure \ref{flux}, the convectively unstable zone is where the convective flux is greater than zero, while the overshooting zone is where the convective flux is smaller than zero. By this criterion, we designed Model NL to have its lower boundary of the convection zone at $0.7123R$, which is consistent with local models and helioseismic results \citep{JCD91,BA97}.

\section{SOUND-SPEED INVERSION}

Helioseismic inversions are based on a linear perturbation analysis of the oscillation equations around a reference model. The differences between the frequencies of the Sun and the reference model are related to the differences in the structure between them.

\begin{figure}
\center
\includegraphics[scale=.7]{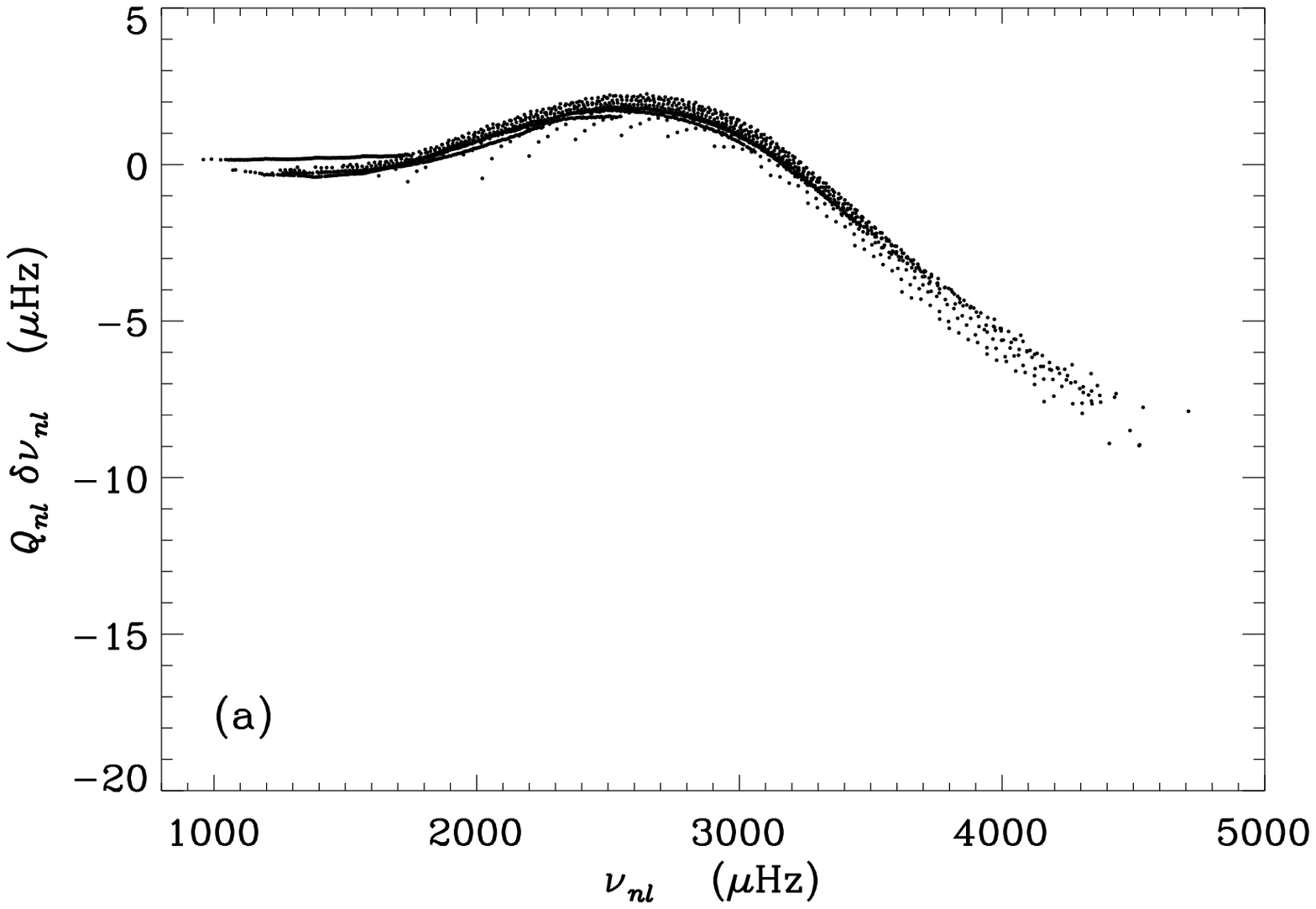}
\includegraphics[scale=.7]{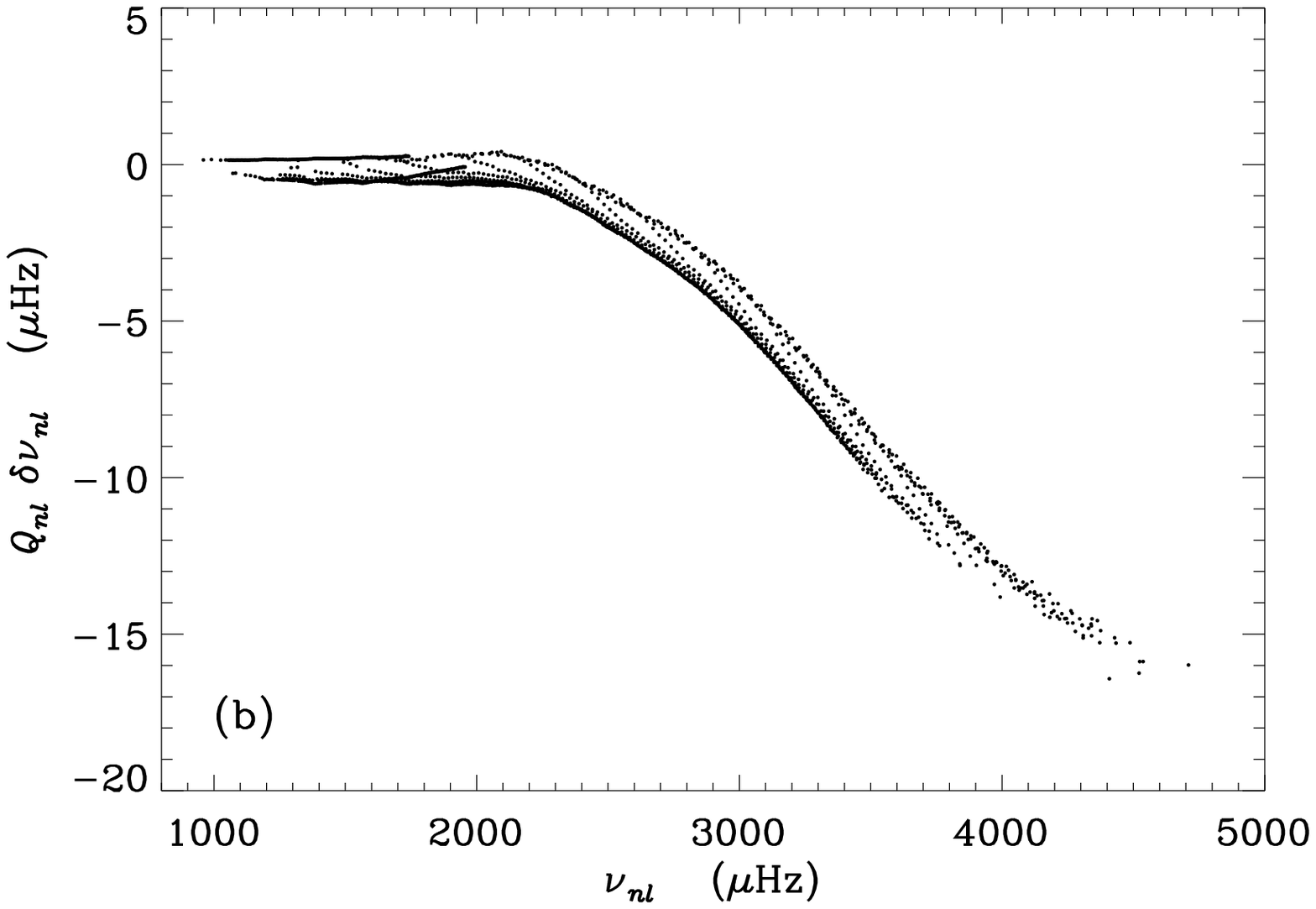}
\caption{Frequency differences between the Sun and solar models, in the sense (Sun)-(Model), scaled by the inertia ratio $Q_{nl}$. Panel (a) shows the scaled frequency differences between the Sun and Model NL for the 1824 modes used for inversion, and panel (b) shows the scaled differences between the Sun and Model S for the same mode set.}
\label{fdiff}
\end{figure}

\begin{figure}
\center
\includegraphics[scale=.7]{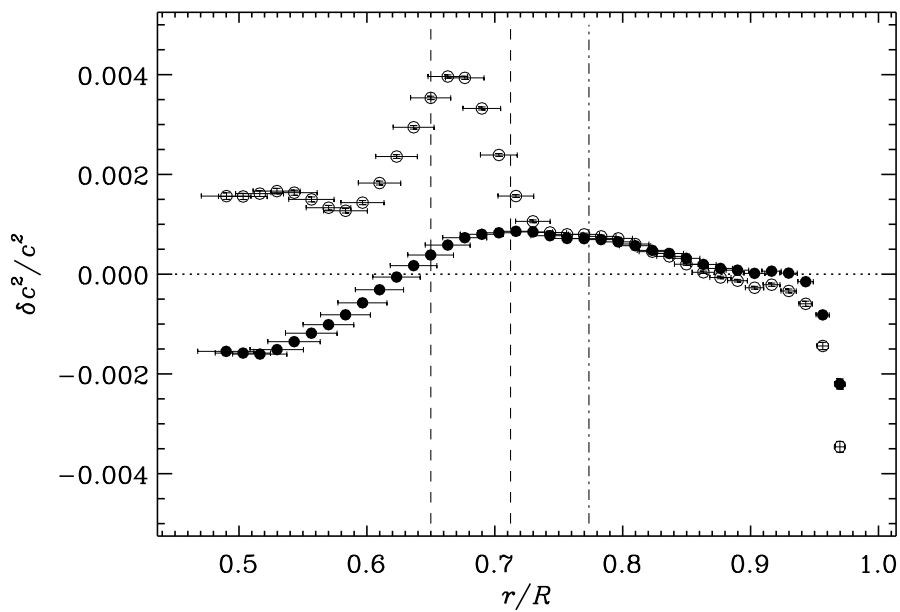}
\caption{Relative differences in squared sound-speed $c^2$ between the Sun and models, in the sense (Sun)-(Model). The open symbols are for Model S, and the filled symbols are for the present Model NL. The vertical error bars correspond to the standard deviations based on the errors in the observed frequencies, whereas the horizontal bars give a measure of the localization of the solution. The dashed lines show the boundaries of the overshooting zone in Model NL, and the dash-dotted line shows where $\nabla = \nabla_{\rm{ad}}$.}
\label{result}
\end{figure}

We use the Aarhus adiabatic oscillation package \citep{ADIPLS} to calculate the oscillation frequencies for both Model NL and Model S. In order to compare with observations, we use data obtained from the MDI experiment on the {\it Solar and Helioseispheric Observatory} spacecraft \citep{SOHO}. Figure \ref{fdiff} shows the scaled frequency differences $Q_{nl}\delta\nu_{nl}$ between the Sun and models, where $Q_{nl}$ is the inertia ratio \citep{JCD96} and $\nu_{nl}$ is the frequency with radial order $n$ and spherical degree $l$. $Q_{nl}\delta\nu_{nl}$ is largely independent of $l$. This reflects that the frequency differences between the Sun and the models are mainly caused by the effects localized near the surface. However, the scaled differences for Model S (Figure \ref{fdiff}(b)) show two distinct branches corresponding to modes that respectively do not, and do, penetrate beyond the localized difference in sound speed between the Sun and this model (cf. Figure \ref{result}).

Since Model NL includes only the envelope above the core, it cannot reproduce the frequencies of the deeply penetrating low-degree modes. In order to avoid the influence of these deeply penetrating modes, we need to make a selection before carrying out the sound-speed inversion to make sure that the lower turning points $r_{\rm t}$ of the modes used in the inversion are well above the bottom of the envelope model. Our criterion is $r_{\rm t}=0.4R$, which leaves 1824 modes from degree $l=10$ up to $l=200$ for inversion.

We then employ the subtractive optimally localized averages method \citep{SOLA} to invert the observed data using Model NL as the reference model. The results are shown in Figure \ref{result} as filled symbols. For comparison, the results of Model S using the same mode set are shown by the open symbols. The well-known bump in sound-speed differences near the bottom of the convection zone ($r\simeq 0.67R$) using Model S is absent in the inversion with respect to Model NL. Since the convection formulation is the only ingredient that differs between the two models, this progress must be due to our improved formulation of convection.

\section{DISCUSSION AND CONCLUSION}

We have shown that the solar envelope model constructed with our nonlocal convection theory has a smooth transition from the convection zone to the radiative interior. The velocity--temperature correlation changes sign crossing the boundaries of the convection zone. Using such a nonlocal convection solar model, the major discrepancy in sound speed between the Sun and SSM can be removed.

Although other ad hoc methods may also be employed to adjust the sound-speed profile, we believe that nonlocal convection is a more physically reasonable solution. Before diffusion was included in solar modeling, the discrepancy in squared sound speed between the Sun and the MLT model was nearly $2\%$ below the convection zone, because the depth of the convection zone in the model was too small \citep{JCD93,Basu10}. Inclusion of diffusion increased the depth of the convection zone but led to a strong composition gradient in the present SSM. To soften this composition gradient, partial mixing in the transition zone was then introduced (e.g., \citealt{RVCD96,BTCZ99,JCD07}). However, in our nonlocal convection theory, there is no abrupt change of physical quantities at the bottom of the convection zone. The nonlocal effects are suitably parameterized, so the solar model has a convection zone of correct depth and a smooth transition region below it \citep{XD01,DX08}.

Moreover, the convective mixing of chemical elements is very efficient in the overshooting zone \citep{DX08}. Here matter is almost fully mixed, therefore there is no abrupt change in composition. Such extended mixing of elements may have important influences on the chemical evolution of the Sun because it transports fragile elements such as lithium to the hot interior where they are destroyed by nuclear reactions \citep{XD09}.

\begin{figure}
\center
\includegraphics[scale=.7]{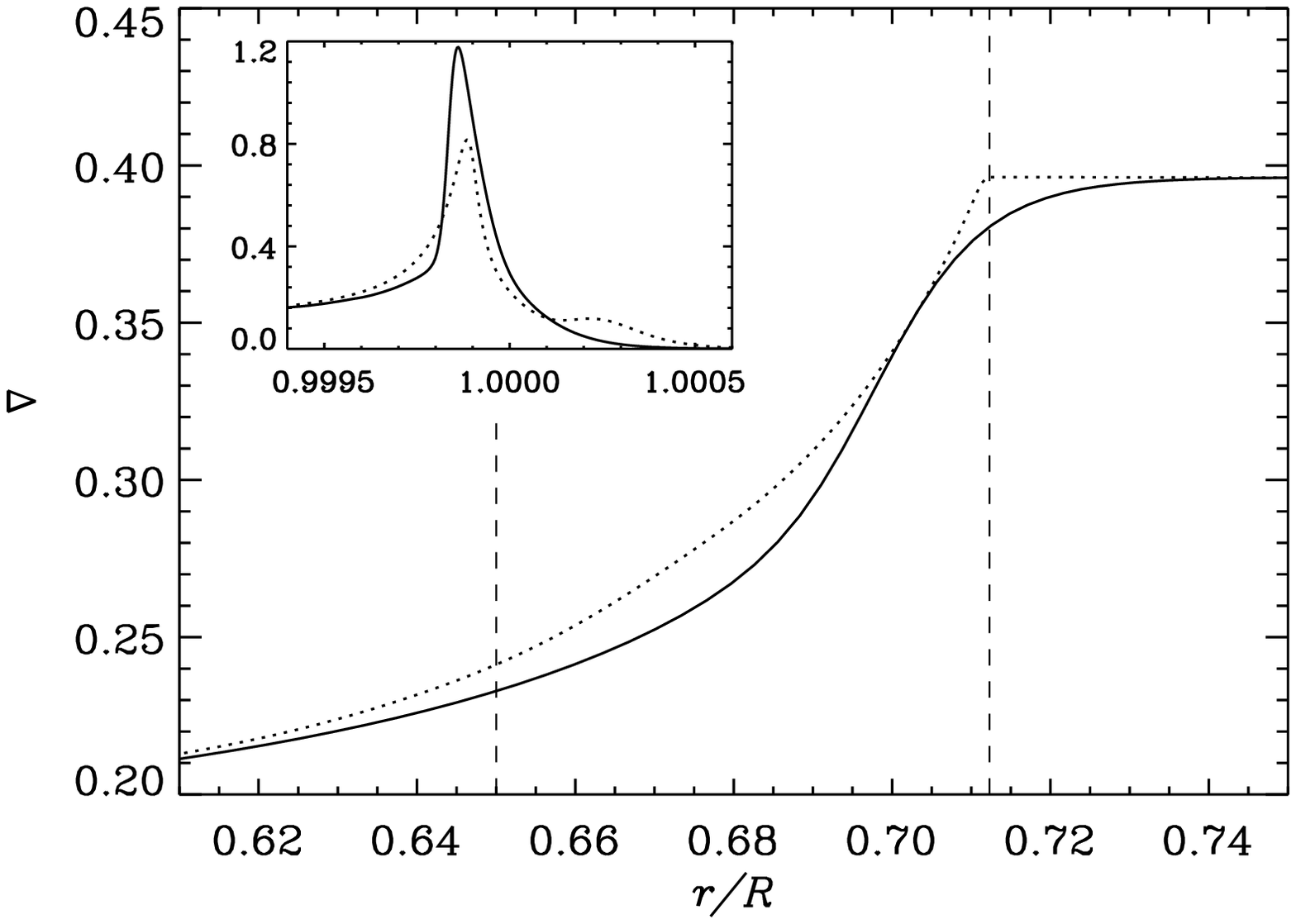}
\caption{Temperature gradient $\nabla$ vs. fractional radius $r/R$ for both Model NL (continuous line) and Model S (dotted line). The dashed lines show the boundaries of the overshooting zone in Model NL. The insert shows $\nabla$ in the near-surface region.}
\label{nabla}
\end{figure}

We note that modifications to the temperature profile resulting from overshooting have also been considered by, for example, \citeauthor{Rempel04} (\citeyear{Rempel04}; for a helioseismic analysis of the result, see \citealt{JCD11}) and \citet{ZL12}; the resulting temperature-gradient profiles were superficially similar to the one obtained here (cf. Figure \ref{nabla}).

Beside solar sound-speed inversion, the nonlocal convection theory also provides a solution to some other long-standing problems in stellar modeling, such as the semi-convection conflicts in massive star evolution \citep{Xiong86}, lithium depletions in late-type dwarfs \citep{XD09}, the theoretical red edge of the classical instability strip, and Mira-like pulsation of red giants \citep{XD07}. Therefore, to get a better understanding of stellar structure, evolution, and oscillation, it is preferable to treat convection nonlocally.

\acknowledgments

This work was supported by the Chinese National Natural Science Foundation (CNNSF) under grants 10773029 and 10973015, and the Ministry of Science and Technology of China under grant 2007CB815406. C. Z. received funding from the exchange program between Chinese Academy of Sciences and the Danish Rectors' Conference (Universities Denmark). Funding for the Stellar Astrophysics Centre is provided by The Danish National Research Foundation. The research is supported by the ASTERISK project (ASTERoseismic Investigations with SONG and Kepler) funded by the European Research Council (grant agreement No.: 267864).

\clearpage

\end{document}